\documentclass{article}

\usepackage{microtype}
\usepackage{graphicx}
\usepackage{subfigure}
\usepackage{booktabs}
\usepackage{aas_macros}

\usepackage{hyperref}

\usepackage[accepted]{icml2024}

\usepackage{amsmath}
\usepackage{amssymb}
\usepackage{mathtools}
\usepackage{amsthm}
\usepackage{physics}
\usepackage{bm}
\usepackage{caption}
\usepackage[capitalize,noabbrev]{cleveref}

\theoremstyle{plain}

\theoremstyle{definition}

\theoremstyle{remark}

\usepackage{xcolor}

\usepackage[textsize=tiny]{todonotes}

\icmltitlerunning{Understanding posterior projection effects with normalizing flows}

\begin{document}

\twocolumn[
\icmltitle{Understanding posterior projection effects with normalizing flows}

\icmlsetsymbol{equal}{*}

\begin{icmlauthorlist}
\icmlauthor{Marco Raveri}{xxx}
\icmlauthor{Cyrille Doux}{yyy}
\icmlauthor{Shivam Pandey}{zzz}
\end{icmlauthorlist}

\icmlaffiliation{xxx}{Department of Physics and INFN, University of Genova, Genova, Italy}
\icmlaffiliation{yyy}{Université Grenoble Alpes, CNRS, LPSC-IN2P3, 38000 Grenoble, France}
\icmlaffiliation{zzz}{Columbia Astrophysics Laboratory, Columbia University, 550 West 120th Street, New York, NY 10027, USA}

\icmlcorrespondingauthor{Marco Raveri}{marco.raveri@unige.it}

\icmlkeywords{Machine Learning, ICML}

\vskip 0.3in
]

\printAffiliationsAndNotice{}  

\begin{abstract}

Many modern applications of Bayesian inference, such as in cosmology, are based on complicated forward models with high-dimensional parameter spaces. This considerably limits the sampling of posterior distributions conditioned on observed data. In turn, this reduces the interpretability of posteriors to their one- and two-dimensional marginal distributions, when more information is available in the full dimensional distributions. We show how to learn smooth and differentiable representations of posterior distributions from their samples using normalizing flows, which we train with an added evidence error loss term, to improve accuracy in multiple ways. Motivated by problems from cosmology, we implement a robust method to obtain one and two-dimensional posterior profiles. These are obtained by optimizing, instead of integrating, over other parameters, and are thus less prone than marginals to so-called projection effects. We also demonstrate how this representation provides an accurate estimator of the Bayesian evidence, with log error at the 0.2 level, allowing accurate model comparison. We test our method on multi-modal mixtures of Gaussians up to dimension 32 before applying it to simulated cosmology examples. 
Our code is publicly available at \href{https://github.com/mraveri/tensiometer}{https://github.com/mraveri/tensiometer}.

\end{abstract}

\section{Introduction}
\label{intro}

While efficient techniques \citep{Lewis:2002ah, 2013PASP..125..306F,2015MNRAS.450L..61H,2009MNRAS.398.1601F,2019OJAp....2E..10F} have been implemented to sample Bayesian posterior distributions up to high dimensions, interpretations are very often limited to one- and two-dimensional marginals distributions, in part due to the finite sample size. 
These distributions are then typically represented as corner plots~\citep{2019arXiv191013970L,corner} and used to derive credible intervals on individual marginalized parameters. 
However, these tools may not be sufficient to reveal the full-dimensional structure of posterior distributions.

The first challenge comes from so-called projection or volume effects, i.e. the apparent distortion of the posterior distribution when marginalizing over subsets of parameters. A typical example of that happens when marginal distributions do not necessarily peak at the maximum-a-posteriori. 
This may occur for non-Gaussian distributions or when poorly constrained, prior-limited, parameters, are projected over significant anisotropic volumes. Such effects may hinder interpretability of models when limited to marginals, for instance multi-probe cosmological analysis \citep{krause2021dark, Joachimi:2021:A&A:}, early dark energy models \citep{2021PhRvD.103f3502M} and galaxy power spectrum analysis \citep{Simon_2023}.
A second challenge is to evaluate the Bayesian evidence, i.e. the normalization of the sampled posterior distribution, that even sophisticated nested samplers sometimes fail at accurately estimating, as shown in \citet{2023MNRAS.521.1184L}. Despite its limitations~\citep{2020MNRAS.496.4647L}, the evidence is a useful metric used for model comparison.

Normalizing flows~\citep{papamakarios2018fast, kingma2017improving, rezende2016variational, papamakarios2018masked,2018arXiv181001367G} have already been used to learn representation of posterior distributions from their samples, allowing, for instance, the efficient computation of tension metrics that do not rely on assuming Gaussian posteriors \citep{Raveri_2021,2022PhRvD.105f3529D}. In this paper, we use state-of-the-art normalizing flow models, with an extra loss term accounting for Bayesian evidence error, to yield accurate Bayesian evidence estimates and to efficiently compute posterior one- and two-dimensional posterior profiles.

Unlike marginal distributions, posterior profiles do not suffer from projection effects as they are essentially insensitive to the volume of the parameter space. Using a simple metaphor, profiling can be thought of as observing the outline of the posterior landscape, whereas marginalization can be seen as measuring its column density. 
As such, they offer a highly complementary tool to analyze posterior and models, which is gaining momentum in fields such as cosmology~\citep[see, e.g.,][]{Karwal:2024qpt,2024arXiv240500261R}.
Note that posterior profiles are well-defined in Bayesian statistics and only require an asymptotic frequentist interpretation when computing parameter constraints.
Profiling, however, has found little application in parametric inference as it requires many optimizations that may only be performed efficiently with differentiable models, such as those provided by normalizing flows. We thus propose an architecture, loss function and training scheme to obtain accurate posterior density estimates and a profiling methodology, all of which are implemented in the TensorFlow Probability frameworks~\citep{2017arXiv171110604D} and publicly available at \href{https://github.com/mraveri/tensiometer}{https://github.com/mraveri/tensiometer}.

The paper is organized as follows: \cref{sec:methods} introduces the formalism of posterior profiles, its difference with marginal distributions, and the normalizing flow architecture used to model posteriors; \cref{sec:related} provides context in the recent literature; \cref{sec:results} presents our benchmark results on analytic multimodal distributions and our applications to cosmology; \cref{sec:discussion} discusses limitations and future work.

\section{Methods}
\label{sec:methods}

\subsection{Relationship between marginalization and profiling} \label{sec:marginalization.vs.profiling}

We first discuss the relationship between marginalization and profiling. To do so, we consider an arbitrary posterior distribution $P$ over a set of parameters partitioned as ${\bm{\theta}=(\bm{\theta}_1,\bm{\theta}_2)}$. The profile posterior for $\bm{\theta}_1$ is obtained by maximizing the joint distribution over $\bm{\theta}_2$:
\begin{align}
    \hat{P} (\bm{\theta}_1) \equiv \max_{\bm{\theta}_2} P(\bm{\theta}_1,\bm{\theta}_2),
\end{align}
and we denote $\hat{\bm{\theta}}_2(\bm{\theta}_1)$ the value of $\bm{\theta}_2$ where this maximum is reached.
If we denote with $P(\bm{\theta}_1)$ the marginal distribution of $\bm{\theta}_1$:
\begin{align} \label{eq:marginal}
    P(\bm{\theta}_1) \equiv \int P(\bm{\theta}_1,\bm{\theta}_2) \, d\, \bm{\theta}_2 \,,
\end{align}
we can write the identity:
\begin{align} \label{eq:marg.profile.1}
\log P(\bm{\theta}_1) = \log \hat{P} (\bm{\theta}_1) +\log \int \frac{P(\bm{\theta}_1, \bm{\theta}_2)}{\hat{P}(\bm{\theta}_1)} \dd{\bm{\theta}_2}.
\end{align}
Since the profile, for all fixed values of $\bm{\theta}_1$ maximizes the joint distribution, \cref{eq:marg.profile.1} shows that:
\begin{align}
\log P(\bm{\theta}_1) - \log \hat{P} (\bm{\theta}_1) & \leq \log \int \mathbb{I}_{P(\bm{\theta}_1,\bm{\theta}_2)>0}  \dd{\bm{\theta}_2} \\ & \leq \log V_2^P(\bm{\theta}_1)
\end{align}
where $V_2^P(\bm{\theta}_1)$ denotes the $\bm{\theta}_2$-volume of the support of the joint distribution $P(\bm{\theta}_1,\bm{\theta}_2)$ at fixed $\bm{\theta}_1$, and $\mathbb{I}$ is the characteristic function.
Note that, in general, $V_2^P(\bm{\theta}_1)$ only depends on boundaries defined by the prior, $\Pi$, such that ${V_2^P(\bm{\theta}_1) \leq V_2^{\Pi}(\bm{\theta}_1)}$.

If we further assume that the posterior distribution is Gaussian, as it is done in~\cite{Hadzhiyska:2023:OJAp:}, we can write with a Taylor expansion:
\begin{align} \label{eq:profile.expansion.1}
& \log P(\bm{\theta}_1) - \log \hat{P} (\bm{\theta}_1) = \nonumber \\
& = +\frac{1}{2}\log\det (\mathsf{\Sigma}_2 - \mathsf{\Sigma}_{21}\mathsf{\Sigma}_1^{-1} \mathsf{\Sigma}_{12}) + \frac{d_2}{2}\log(2\pi) + \dots \nonumber \\
&= -\frac{1}{2}\log\det \mathsf{F}^{(2)}_P(\bm{\theta}_1) + \frac{d_2}{2}\log(2\pi) + \dots,
\end{align}
where $\mathsf{\Sigma}_\cdot$ denotes the blocks of the partitioned covariance matrix, $d_2$ the dimension of the $\bm{\theta}_2$ subspace, and where we have defined
\begin{align} \label{eq:fisher.matrix.2}
\mathsf{F}^{(2)}_P(\bm{\theta}_1) = \eval{\frac{1}{2}\pdv[2]{\log P(\bm{\theta}_1,\bm{\theta}_2)}{\bm{\theta}_2}}_{\bm{\theta}_1,\hat{\bm{\theta}}_2(\bm{\theta}_1)},
\end{align}
as the empirical Fisher matrix of $\bm{\theta}_2$, at fixed $\bm{\theta}_1$, and optimum $\hat{\bm{\theta}}_2(\bm{\theta}_1)$.
Note that $\mathsf{F}^{(2)}_P$ is the empirical Fisher matrix -- the second derivative of the log posterior of the observed data realization -- and not the real Fisher matrix, as it is not averaged over data realization and includes the prior.
\Cref{eq:profile.expansion.1} can be seen as the first order term of a Laplace expansion of the difference between the marginal and profile distributions.

\begin{figure}[t!]
\includegraphics[width=\columnwidth]{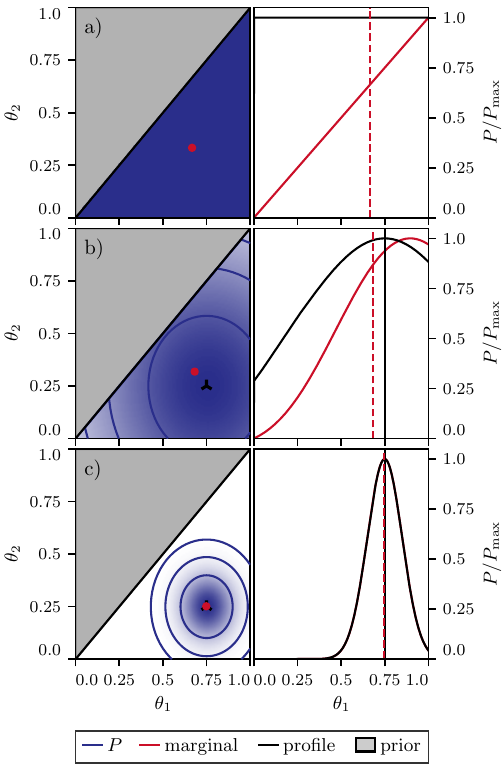}
\caption{\label{fig:example.Gaussian}
{\bf Gaussian example:} in the left column, the joint prior (gray) and posterior (blue) distributions for different Gaussian distributions, with increasingly strong constraints from top to bottom. The right column shows the marginal distribution of the posterior over $\theta_1$ (red) and the corresponding profile (black). The difference between the two curves decreases as the posterior is better constrained.
Note that this example is symmetric if we were to exchange $\theta_1$ and $\theta_2$.
}
\end{figure}

Decomposing the empirical Fisher matrix into its likelihood ($\mathcal{L}$) and prior ($\Pi$) components, we can also write the first order term as
\begin{align} \label{eq:projection.effects}
& \log P(\bm{\theta}_1) - \log \hat{P} (\bm{\theta}_1) \nonumber \\
& \leq -\frac{1}{2}\log\det \mathsf{F}^{(2)}_{\mathcal{L}} -\frac{1}{2}\log\det \mathsf{F}^{(2)}_{\Pi} + \frac{d_2}{2}\log(2\pi) \nonumber \\
& = -\frac{1}{2}\log\det \mathsf{F}^{(2)}_{\mathcal{L}}(\bm{\theta}_1) +\log V_{\Pi}^{(2)}(\bm{\theta}_1)  \\
& \leq \log V_2^P(\bm{\theta}_1) \nonumber
\end{align}
This clarifies why projection effects arise and are known with two different names: projection or volume effects.
If, in \cref{eq:projection.effects}, the likelihood term dominates the leading order discrepancy between profile and marginal distributions, then this means that the empirical Fisher matrix depends on position, which is due to non-Gaussianities of the likelihood, and corresponds to what is colloquially referred to as a projection effect~\citep{2016MNRAS.456L.132S}.
If, instead, the prior term dominates, this means that the data is weakly constraining, and marginalization suffers from prior volume effects, i.e. differences in prior volume along the parameter line of sight. These two cases are illustrated in \cref{fig:example.Gaussian} with a two-dimensional triangular prior and increasingly tighter posterior distributions.

\subsection{Normalizing flow architecture, loss and training}

Most performance metrics applied to generative models have to do with marginal distributions -- potentially of derived parameters or quantities computed from generated samples~\citep{2023JPhCS2438a2155R,2023arXiv230212024C}. However, for the problem at hand, we are rather concerned with the accuracy of the (logarithm of the) posterior probability density function.

\subsubsection{Architecture}

To tackle this challenge, we tested various architectures that combine two of the best-performing types of normalizing flows (NF), namely Masked Autoregressive Flows~\citep[MAF,][]{papamakarios2018masked} and Neural Spline Flows \citep{durkan2019neural}.  In the best performing architecture, we stack a number of MAFs that implement autoregressive affine transformations, each parametrized by a unique neural network (with masked inputs). We also insert random parameter permutations between each MAF. In addition, we select sequences of permutations with a low variance between parameter coordinates to maximize the mixing of coordinates throughout the sequence of autoregressive transformations. This architecture provides light-weight flexibility allowing to model posterior distributions to high accuracy. We found that adding spline flows or replacing MAFs by spline flows resulted, on our finite samples, in overfitting and/or noisier posterior profiles (see \cref{sec:discussion}).

\subsubsection{Evidence error loss function}

To further improve the accuracy of the estimated posterior density, we propose to add a term to the standard normalizing flow loss. However, the posterior density is usually computed from the product of the likelihood and priors, and it normalization, given by the Bayesian evidence, is a priori unknown. Nevertheless, this normalization constant is the same for all available samples, motivating a loss function that reduces the scatter of the approximate density around an unknown mean. Denoting $\log q(\bm{\theta})$ the approximate, normalized flow density, and $\log \tilde{P}(\bm{\theta})$ the unnormalized posterior density, we define the evidence error loss (EEL) for a batch of samples $\qty{\bm{\theta}_i}_{1\leq i < N}$ drawn from the posterior as
\begin{equation}\label{eq:EEL}
    {\rm EEL}(\qty{\bm{\theta}_i}) \equiv \frac{1}{N}\sum_{i=1}^N \qty(\log \frac{q(\bm{\theta}_i)}{\tilde{P}(\bm{\theta}_i)} - \log\hat{\mathcal{Z}}(\qty{\bm{\theta}_i}))^2,
\end{equation}
where
\begin{equation}
    \log\hat{\mathcal{Z}}(\qty{\bm{\theta}_i}) \equiv \frac{1}{N} \sum_{i=1}^N \log \frac{q(\bm{\theta}_i)}{\tilde{P}(\bm{\theta}_i)}
    \label{eq:evidence_estimator}
\end{equation}
is an estimator of the (logarithm of the) evidence.
We add this term to the standard Kullback-Leibler divergence NF loss, given by ${\sum_{i=1}^N \log q(\bm{\theta}_i) / N}$, and use a soft adaptation scheme \citep{heydari2019softadapt} to balance the two terms during training.
Once trained, we use \cref{eq:evidence_estimator} to estimate the evidence over the full posterior training sample. Similar evidence estimator and loss functions were also recently suggested by \citet{Polanska_2024,Srinivasan_2024}.

\subsubsection{Training}

In addition to the specific architecture and evidence error loss term, we obtain more accurate and stable results by training populations of normalizing flows and averaging the individual density estimates~\citep{2016arXiv161201474L,2019MNRAS.488.4440A}.

At last, we implement a new and simple adaptive learning rate modulation scheme as follows. At every epoch end, we fit a line through the validation loss evaluated over the last $N_{\rm epochs}$ epochs (by default, 25). If the slope is negative (as expected during learning), the learning rate is unchanged; if it is positive, then we multiply the learning rate by a factor $\alpha$ (by default, $1/\sqrt{10}$). We train all our models with an initial learning rate of $10^{-2}$ and stop training when the learning rate reaches $10^{-5}$.

\begin{figure*}[ht!]
\centering
\includegraphics[width=\columnwidth]{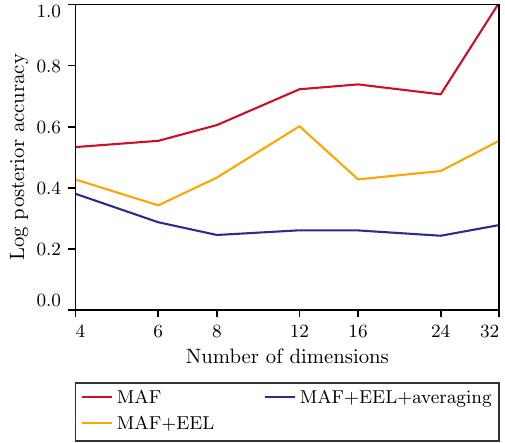}
\includegraphics[width=\columnwidth]{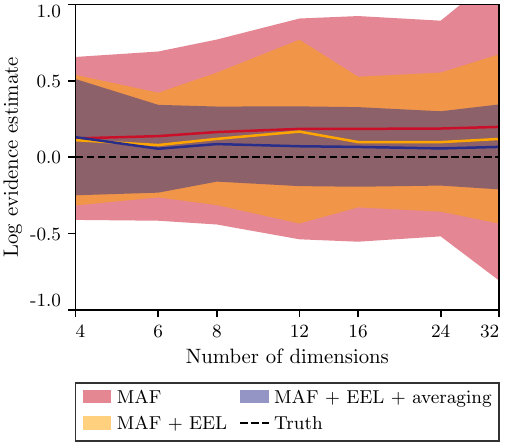}
\caption{ \label{fig:mg_evidence_error}
{\bf Mixture of Gaussian example:} 
Evidence accuracy (left) and error (right) estimated from different flow architectures. MAF only uses a single standard masked auto-regressive flow, MAF+EEL additionally adds the evidence error term (Eq.~\ref{eq:EEL}) to the loss function and MAF+EEL+averaging additionally averages over six flows. We see that including these additional modifications to the architecture significantly improves the accuracy and constraints on the evidence. }
\end{figure*}

\subsection{Profiling}

Once a population of normalizing flows has been trained, we aim at deriving one- and two-dimensional posterior profiles. To do so, we define, for each parameter of interest $\theta_i$ a binning (typically 64 linearly spaced bins), and we optimize over other parameters, $\bm{\theta}_{j \neq i}$, to evaluate the profile $\hat{P}(\theta_i)$. Our algorithm works as follows:
\begin{enumerate}
    \item We sample the NF population (which is fast) sufficiently many times such that each bin in $\theta_i$ contains at least one sample. 
    \item Within each bin, we save the sample with the highest density $q$, which already provides a noisy estimate of the profile. Note here that the value of $\theta_i$ is not necessarily located at the center of the bin.
    \item We improve this estimate by optimizing, in each $\theta_i$ bin, the flow density $q$ over other parameters $\bm{\theta}_{j \neq i}$ at fixed $\theta_i$ value (the value of the initial sample in the bin), and denote $\hat{q}(\theta_i)$ the optimum of the flow density. This optimization is performed using gradient descent, which is vectorized in our code (i.e. all bins are simultaneously optimized), making it reasonably fast.
    \item The final profile is estimated by linearly interpolating between the values of $(\theta_i,\hat{q}(\theta_i))$.
\end{enumerate}
This algorithms is easily generalized to the two-dimensional case, with two-dimensional bins over $(\theta_i,\theta_j)$ and optimizations over $\bm{\theta}_{k\neq i,j}$. While this is in general prohibitively expensive due to the large number of optimizations in high-dimensional spaces, the NF flow density can be evaluated efficiently, making it possible to obtain stable profiles within minutes.

Finally, we make it possible to obtain profiles over derived parameters, i.e. parameters that can be computed from the parameters sampled during posterior inference. To do so, we transform the flow density according to the reparametrization, using either analytic TensorFlow Probability bijectors when the mapping is simple, or by training an additional normalizing flow to learn it. In particular, we apply this latter functionality to obtain profiles over the effective cosmic structure parameter $\sigma_8$, which is computed by the theoretical model from more fundamental parameters which are themselves used during posterior inference (see \cref{sec:Cosmo.Applications}). Note that one may not, in general, simply train a flow on derived parameters, as the posterior density $P$ used in the EEL loss shown in \cref{eq:EEL} corresponds to a specific choice of parameters.

\begin{figure*}[h!]
    \centering
    \includegraphics[width=\columnwidth, height=0.7\textwidth]{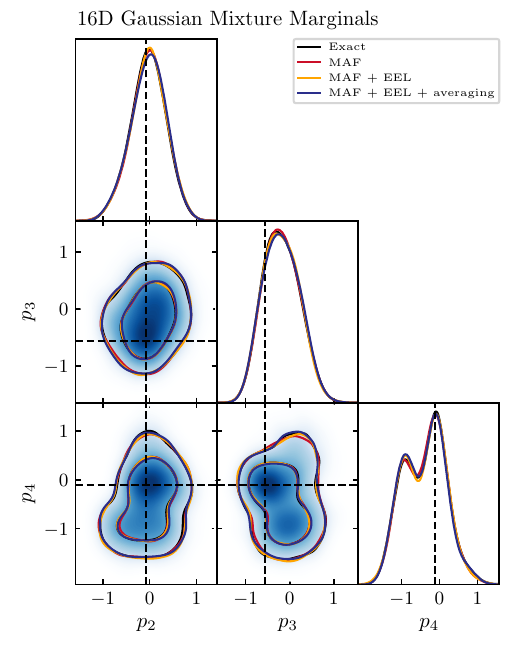}
    \includegraphics[width=\columnwidth, height=0.7\textwidth]{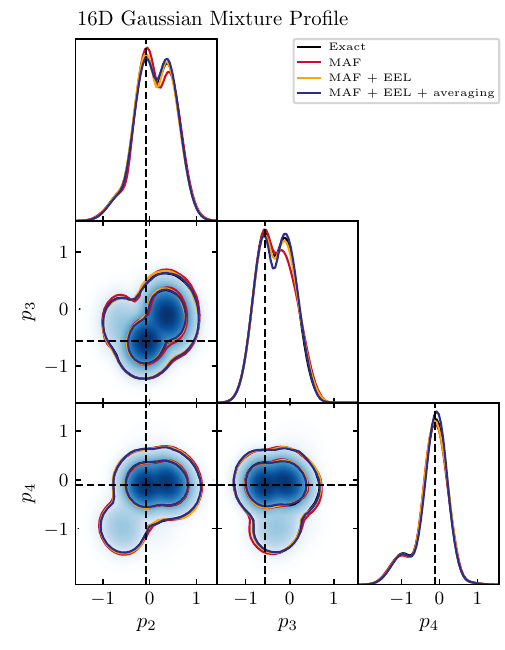}
    \caption{{\label{fig:mg_marginal_profile} \bf Mixture of Gaussian example:} Marginalized (left) and profile (right) posteriors of three parameters in the 16-dimensional mixture of Gaussians. We see that while all three models perform well in terms of capturing the marginal posterior, using just a standard MAF struggles to capture the profiles posterior of some parameters. Adding EEL and averaging significantly helps in obtaining unbiased marginal and posterior distributions.}
\end{figure*}

\begin{figure}
\includegraphics[width=\columnwidth]{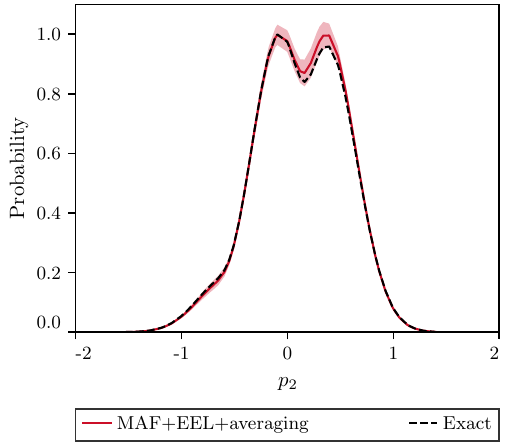}
\caption{ \label{fig:mg_profile_error}
{\bf Mixture of Gaussian example:} 
the red band shows the mean profile estimated from averaging six flows, as well as its uncertainty computed from the variance of the profiles estimated for each flow individually. The exact profile, calculated from maximizing the analytic density, is consistent with the flow profile, within one standard deviation.}
\end{figure}

\section{Related works}
\label{sec:related}

Several previous studies have developed methods based on machine-learning to learn the posterior distribution, but most of them focus on obtaining accurate marginal densities~\citep{radev2021amortized, Raveri_2021} or evidence estimates \cite{turner2014generalized, Srinivasan_2024,Polanska_2024, Jeffrey_2024}. However, we additionally aim to obtain reliable profiles of the parameters, which requires maximization in a high-dimensional parameter space. This, in turn, requires accurate estimates of posterior density values in the high-dimensional spaces (the full dimension minus one or two). For this purpose, we combine the standard NF loss with the evidence error loss, which significantly improves the posterior density values. 

While preparing this manuscript, \citet{Srinivasan_2024} published a study that adds a similar term to their loss function, resulting in better estimates of evidence. However, compared to them, we use a stack of normalizing flows to obtain a flexible and easy to optimize model. Moreover, we train and average multiple flows to reduce the noise in the estimate of posterior in the high dimensional space. Finally, we develop a GPU-supported optimization routine to estimate the parameter profile for large number of parameter values. All these improvements are crucial to obtain reliable parameter profiles, especially in two-dimensional corner plots due to large number of bin combinations to estimate the profile maximization. In addition to applying this architecture to analytic examples, such as mixture of Gaussians, we also apply it to cosmological setting, obtaining reliable posterior profiles. Finally, we add the functionality to obtain profiles on the derived parameters (non-linear combination of sampled parameters) which is generally useful in cosmological examples.

\section{Results}
\label{sec:results}

\subsection{Benchmark with mixtures of Gaussians}

To test the performance of our model and profiling algorithm, we generate test datasets of multi-dimensional mixtures of Gaussian of varying dimensions $d$. The probability distribution function is given by:
\begin{equation}
    P_{\rm MG}(\bm{\theta}) = \sum_i^{n_g} w_i \, \mathcal{N}(\bm{\theta}|\bm{\mu}_i, \mathsf{\Sigma}_i),
    \label{eq:MG_density}
\end{equation}
where, $n_g$ is the number of Gaussian components, $w_i$ is the weight of $i$-th Gaussian such that $\sum_i^{n_g} w_i = 1$, and $\bm{\mu}_i$ and $\mathsf{\Sigma}_i$ are the mean and covariance of $i$-th Gaussian. We fix ${n_g = 5}$ and generate random $\bm{\mu}_i$ in the range (-1, 1) for each dimension, and $\mathsf{\Sigma}_i$ is a random non-diagonal covariance matrix for each Gaussian component, drawn such that the mixture exhibits multimodality. Since this is a normalized sum of Gaussians, the evidence of this distribution is ${\log \mathcal{Z}_{\rm MG} = 0}$. 

We then propose a specific architecture: each flow is composed of ${\lceil 2\log_2{d} \rceil + 2}$ MAFs, each parametrized by masked neural networks~\citep{2015arXiv150203509G} with two hidden layers of $2d$~features and using asinh activation functions. In our final setup, we train six such flows that are averaged.

We first quantify the performances of our architecture and training methodology by measuring the evidence accuracy and its uncertainty (\cref{fig:mg_evidence_error}) on examples with dimensions varying from 4 to 32. The evidence uncertainty is computed from the standard deviation of the evidence estimate over all samples (see \cref{eq:evidence_estimator}). We first train flows composed of a sequence of MAFs with the standard loss function, showing that the evidence uncertainty cannot go below 0.6 for mixtures of Gaussians for $d>8$, as shown on the left panel of \cref{fig:mg_evidence_error} (MAF). We then demonstrate that when adding the evidence error loss (EEL) term (\cref{eq:EEL}), this uncertainty reaches 0.4-0.5, while preserving the quality of marginal distributions (MAF+EEL). We then train populations of flows and average their density estimates, which allows us to reach evidence uncertainties at the 0.2 level (MAF+EEL+averaging). Following the same procedure, we show in the right panel of \cref{fig:mg_evidence_error} that adding the EEL loss and averaging multiple flows allows us to reduce the bias on the evidence estimate, while noting it remains within uncertainties for all tested configurations.

In \cref{fig:mg_marginal_profile}, we show the performance of our model and profiling algorithms in ${d=16}$ dimensions. In the left panel of \cref{fig:mg_marginal_profile}, we compare the performance of the models in capturing the marginal posterior distribution for 3 of the 16 parameters in this space, finding that even a single standard MAF captures the marginal distribution well. In the right panel, we compare the profile posterior distribution for the same three parameters. In this analytic example, we can obtain the exact profiles by maximizing the probability density of \cref{eq:MG_density}, even in ${d=16}$ dimensions, as shown by black solid lines. We find that the posterior profiles differ significantly from the posterior marginal distributions, due to the multimodality of this example. We find that, in this case, a single standard MAF fails to capture the profiles of parameters $p_2$ and $p_3$ accurately, whereas training with the extra EEL term and averaging allows us to correctly capture this multi-modality. We also estimate the two-dimensional profile distributions, which requires optimizations in 14 dimensions for all the $32\times32$ two-dimensional bins, for each pair of parameters. Note that this optimization would be computationally prohibitive without a differentiable estimate of the posterior density as we have developed here. We find similar performances in the one- and two-dimensional profiles when training the flow with EEL term and averaging for various dimensions up to 32.

Finally, in \cref{fig:mg_profile_error}, we use profiles derived from each of the individual flows to estimate the variance of the profile estimated from the averaged flows, showing the result for one parameter. We see that the averaged flow profile lies within one standard deviation of the exact profile. 

\subsection{Application to cosmological simulated data}
\label{sec:Cosmo.Applications}

\begin{figure}
\includegraphics[width=\columnwidth]{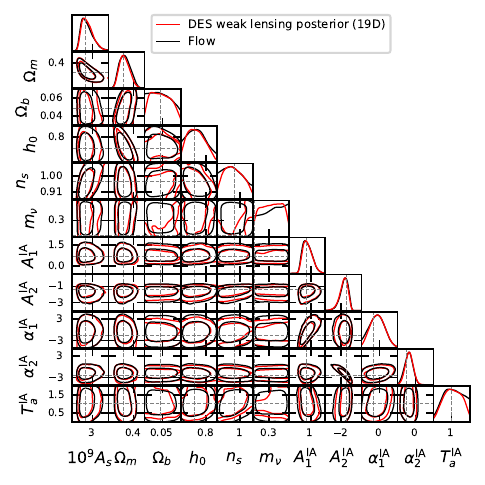}
\caption{ \label{fig:des_marginals} 
{\bf Cosmology example:} 
We show the marginalized distribution of posterior for various parameter combinations as estimate from the true chain and from normalizing flow (MAF+EEL+averaging). We see that the flow can capture various non-Gaussian features accurately.}
\end{figure}

\begin{figure*}[h!]
\centering
\includegraphics[width=\columnwidth]{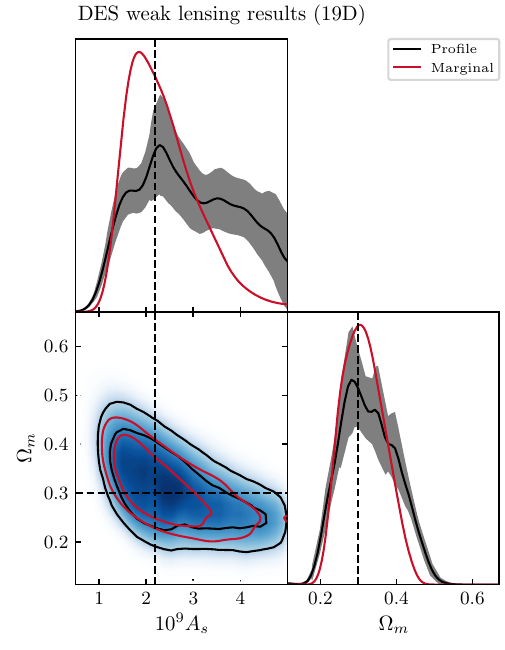}
\includegraphics[width=\columnwidth]{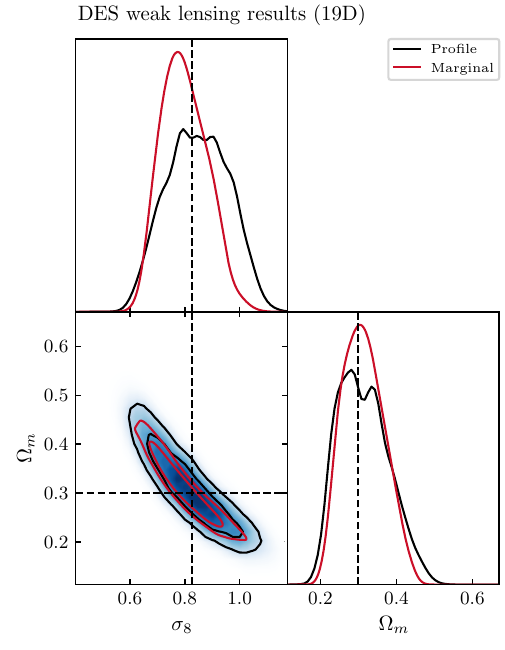}
\caption{ \label{fig:des_profiles}
{\bf Cosmology example:} 
\textit{Left}: We show the application of our flow architecture to estimate the profile and marginals of two cosmological parameters within a 19 dimensional parameter space. 
\textit{Right}: We additionally show the constraints on a derived parameter, $\sigma_8$, which is inferred from other cosmological parameters. In this case, the mapping from the original sampling space is learned as well.
In both panels the blue shade shows the 2D profile while the contours show the 68\% and 95\% integral regions of the profile and marginal distributions respectively.
}
\end{figure*}

Given the good performance of our model and profiling algorithm for multimodal, high-dimensional mixtures of Gaussian, we now apply it to the analysis of simulated cosmological data.

Specifically, we analyze simulated two-point correlation functions constructed out of galaxy positions and shapes. This is the statistics of choice of astronomical imaging surveys to probe the distribution of matter in the Universe, and constrain cosmological models of dark energy~\citep{Mandelbaum_2018}. To do so, we use the theoretical model described in \citet{krause2021dark} to create a mock data vector (dimension 400) emulating weak lensing data from the Dark Energy Survey ~\citep[see Fig. 5 of][]{Abbott_2022}. The model has 19 parameters, which includes six cosmological parameters as well as five astrophysical and eight observational systematics parameters.
Using the same theoretical model and the covariance matrix of \citet{krause2021dark}, we then sample the posterior distribution for this simulated data vector. To do so, we use a nested sampling algorithm, PolyChord~\citep{2015MNRAS.450L..61H}, set to high accuracy to obtain 0.5~million samples. While this is highly sufficient to estimate the one- and two-dimensional marginal posterior distributions, it is inadequate for profiling. Finally, we use this sample to train the normalizing flow architecture described above and learn a smooth representation of the 19-dimensional posterior. \Cref{fig:des_marginals} shows the marginal distributions for cosmological and astrophysical parameters derived from the PolyChord sample and the trained flow. This figure illustrates the non-Gaussianity of this posterior, which arises from both the non-linearity of the theoretical model and prior volume effects, all well-captured by the flow.

In \cref{fig:des_profiles}, we compare the marginal distributions and profiles for two cosmological parameters of interest, out of 19. We show the constraints on the matter density of the Universe ($\Omega_m$) and the amplitude of primordial fluctuations $A_s$ on the left. On the right, we additionally show the profiles of $\Omega_m$ and a derived parameter, $\sigma_8$, which measures the amplitude of late time matter density fluctuations, and is computed from the theoretical model as a non-linear function of the other cosmological parameters. We also show the true value used to create the mock data vector with dashed lines (the maximum a-posteriori, MAP value). We see that, as expected, the marginalized posteriors peak at different values compared to the true MAP value, whereas the profile posteriors peak at the true values. Additionally, the left panel shows that the parameter $A_s$ is, according to its profile, only weakly constrained by the data, unlike what is suggested by the peaked marginal distribution.

\section{Discussion and future work}
\label{sec:discussion}

One- and two-dimensional marginal distributions are often insufficient to capture the properties of high-dimensional posterior distributions derived from complex data and model. These most notably suffer from so-called projection effects, due to integrating the posterior along non-Gaussian or unconstrained parameter directions, unlike posterior profiles, obtained my maximization. However, obtaining posterior profiles requires an accurate estimate of the actual posterior values, which standard normalizing flows fail to capture. In this paper, we thus provide a normalizing flow architecture and simultaneously minimize the standard Kullback-Leibler divergence loss and an extra term, the evidence error loss, which dramatically improves the quality of posterior profiles, and provides an estimate of the Bayesian evidence as a by-product. We validate our method on analytic examples using mixtures of Gaussians up to dimension 32, and then apply it to a simulated data analysis from cosmology.
We note here that obtaining stable profiles also required training an ensemble of flows using an adaptive loss weighting scheme and a new adaptive learning rate scheduler. 

In the future, we aim to improve the accuracy and stability of the profiles. In particular, neural spline flows offer a promising avenue in terms of flexibility. However, our tests with spline flows resulted in overfitting, likely due to the large yet finite size of our posterior samples, and inaccurate profiles. We could curtail this effect by constraining the underlying spline parameters, thus limiting fluctuations in the Jacobian entering the log density evaluation. Another promising avenue, though, consists in merging the posterior sampling and flow training steps using an iterative method, as suggested by multiple studies~\citep{pmlr-v37-rezende15,2020arXiv200105486G,2022PNAS..11909420G,2022arXiv220514240G,2022JOSS....7.4634K,2023JOSS....8.5021W}. This would alleviate accuracy issues related to limited samples, and our profiling methodology could be readily applied. The code is publicly available at  \url{https://github.com/mraveri/tensiometer}

\section*{Acknowledgements}
We thank 
Tanvi Karwal,
Junpeng Lao and
Minsu Park
for the helpful discussions and comments.
Computing resources were provided by the National Energy Research Scientific Computing Center (NERSC), a U.S. Department of Energy Office of Science User Facility operated under Contract No. DE-AC02-05CH11231, and by the University of Chicago Research Computing Center through the Kavli Institute for Cosmological Physics. 
M.R. acknowledges financial support from the INFN InDark initiative and from the ICSC Spoke~2 under the grant NFL-GW. 
M.R. and C.D. acknowledge financial support from the Centre de Physique Théorique de Grenoble-Alpes. SP is supported by the Simons Collaboration on Learning the Universe.

\bibliography{icml_paper}
\bibliographystyle{icml2024}

\end{document}